\documentstyle[pra,aps,twocolumn,psfig]{revtex}
\begin{document}
\title{Thermodynamic properties
of confined interacting Bose gases -
a renormalization group approach}
\author{Gernot~Alber} 
\address{Abteilung f\"ur Quantenphysik, Universit\"at Ulm,
D--89069 Ulm, Germany\\
(submitted to Phys. Rev. A)
}
\maketitle
\begin{abstract}
A renormalization group method is developed with which
thermodynamic properties of
a weakly interacting, confined Bose gas can be investigated.
Thereby effects originating from a confining potential
are taken into account by periodic boundary conditions 
and by treating the resulting discrete energy levels
of the confined degrees of freedom properly.
The resulting density of states
modifies the flow equations of the renormalization group
in momentum space. It is shown that as soon
as the characteristic length of confinement becomes comparable
to the thermal wave length of a weakly interacting and trapped
Bose gas its
thermodynamic properties are changed significantly.
This is exemplified by investigating characteristic bunching properties
of the interacting Bose gas
which manifest themselves in the second order coherence
factor.
\end{abstract}
\narrowtext

\section{Introduction}

The experimental realization of Bose Einstein condensation
of trapped, ultra cold, weakly interacting atomic gases
\cite{Boulder,Ketterle}
has renewed the interest in their thermodynamic properties.
Definitely,
many of these properties have already been studied
more than  thirty years ago 
in connection with the theoretical efforts to describe
superfluidity
of strongly interacting helium atoms. 
However, in these recent experimental realizations the
physics of the trapping process
and of the internal structure of the trapped bosons
still pose interesting new theoretical questions which
are not yet understood completely \cite{Stringari}.

So far the majority of theoretical approaches aiming at
describing thermodynamic properties of weakly interacting,
trapped
Bose gases has concentrated on mean field approaches \cite{Stringari}.
At zero temperature these mean field approaches lead to the
well known Gross-Pitaevski equation and to their multicomponent
generalizations. These theoretical descriptions 
have proven useful in obtaining  first quantitative understandings of
many  recent trapping experiments \cite{Ketterle}.
However, it is known that mean field theories do not yield an
accurate description of thermodynamic properties of interacting
gases close to the transition point of a phase transition \cite{Kerson}.
Furthermore, with improving experimental accuracy it is expected
that in the near future possible
deviations from mean field results might become
accessible experimentally even in regimes far away from transition
points.

Motivated by these prospects in this paper
a theoretical description
of thermodynamic properties of confined, 
weakly interacting Bose gases
is developed which is based on renormalization group (RG) methods
\cite{Wilson,Fisher,Wiegel}.
RG methods are known to yield accurate descriptions of thermodynamic
partition functions also in regions close
to a second order phase transition.
Furthermore, by these methods
certain universal quantities, such as 
critical exponents, might even be evaluated by analytical means.
Typically the evaluation of thermodynamic
partition functions with the help of
RG methods is only slightly more complicated than 
by mean field methods. Thus, these RG methods represent 
a convenient theoretical approach complementing
elaborate numerical Monte-Carlo simulations of partition functions 
with which one may also test limitations of mean field theories.

In the special case of an unconfined,
homogeneous three-dimensional interacting
Bose gas such a RG approach 
has already been developed recently by Bijlsma and Stoof \cite{Stoof}. 
However, their quantitative predictions of the dependence
of the transition temperature on the scattering length, for example,
have been more than one order
of magnitude larger than corresponding numerical predictions
which are based on elaborate Monte-Carlo simulations of the
relevant partition function \cite{grueter}.
In a recent perturbative approach this discrepancy has been clarified
\cite{Holzmann1,Holzmann2}. These perturbative
studies are in satisfactory agreement with
the RG calculations of Bijlsma and Stoof \cite{Stoof} and with
recent experiments \cite{transition} and attribute
the numerical inaccuracies of the previous
Monte-Carlo approach to incorrect
extrapolation procedures.
This example demonstrates that despite
extensive previous work 
there are still open 
questions even in the context of the traditional
case of a homogeneous and unconfined interacting Bose gas \cite{Huang2}.
In addition, the experimental
possibility to control and manipulate trapped Bose condensates
has raised many new interesting physical questions most of which
have not been tackled yet theoretically beyond the framework of 
mean field approximations. 

In this paper a RG approach is developed which is capable
of determining thermodynamic properties of weakly interacting
and confined Bose gases.  This approach aims at describing
physical effects of spatial confinement beyond the local density
approximation by modeling
effects originating from confinement
by periodic boundary conditions.
However, contrary to the usual continuum-limit-approximation
within our approach the required summations over all possible
states of the atomic center of mass motion are not replaced 
by integrations but the discrete energy levels resulting from the
confined degrees of freedom are taken into account.
Thus the presented approach should yield a valid description
even in cases
in which the thermal wave length of an interacting Bose gas
becomes comparable to the characteristic length of confinement
and in which the local density approximation breaks down.
Indeed in the subsequent treatment it
is demonstrated that as soon as the confinement length becomes
comparable to the thermal wave length of the interacting Bose gas
thermodynamic properties of an interacting Bose gas
are changed significantly. This
characteristic behavior is exemplified by investigating
various thermodynamic properties of an interacting
Bose gas, such as
its bunching properties 
which manifest themselves in the second order coherence factor
and which
are accessible to experimental observation.
Though the presented RG approach differs slightly from the one
developed by Bijlsma and Stoof \cite{Stoof} previously 
it is demonstrated that resulting critical properties, such as
the dependence of the transition temperature on the scattering
length,
are in satisfactory agreement with these previous RG results and
with the recent perturbative calculations of Holzmann et al.
\cite{Holzmann1,Holzmann2}.

The paper is organized as follows:
In Sec.II the general theoretical approach is developed.
Starting from the path integral representation of the grand canonical
partition function an RG method is developed for its evaluation.
As effects of a confining potential are taken into account
by periodic boundary conditions which still guarantee
translational invariance of the problem
the RG procedure is developed in momentum space. As a main
approximation the random phase approximation (RPA) \cite{Hertz}
is used which is applied to the symmetry broken phase.
In Sec.III resulting numerical results are presented.
Firstly, we concentrate on characteristic properties of 
an unconfined homogeneous three-dimensional Bose gas which have
been of interest recently. Secondly, the influence of
confinement is exemplified by investigating isothermal properties
of the pressure and of the second order coherence
factor of an interacting Bose gas. 


\section{Theoretical treatment}
In this section a RG approach is developed for the evaluation
of the grand canonical partition function of an interacting and
confined Bose gas in $D$ spatial dimensions.
Effects of confinement are taken into account
by periodic boundary conditions and by proper summations over
the discrete energies associated with the confined degrees of freedom.
Accordingly,
the RG approach is developed in momentum space. 
The RG flow equations are derived for the temperature
regime below the phase transition with the help of the random
phase approximation (RPA) \cite{Hertz}. 
This dynamical regime is characterized by a non-zero order parameter.
Details of the derivation of these RG equations are given in
appendix A.
For the sake of completeness
the RG equations which apply
to the dynamical region above the phase transition, i.e.
in the region of a vanishing order parameter,
are summarized
in appendix B.
These latter equations
been already discussed in detail previously \cite{Stoof,Hertz}. 

\subsection{Path integral representation of the partition function}
Thermodynamic properties of an interacting system of bosons can
be described by its (grand-canonical) partition function \cite{Feynman}
\begin{eqnarray}
Z(z,\beta) &=& {\rm Tr} [e^{-\beta({\bf H} - \mu {\bf N})}]
\label{partition}
\end{eqnarray}
with the fugacity $z=e^{\beta\mu}$
($\mu$ denotes the chemical potential) and with
the inverse temperature
$\beta = 1/(kT)$ ($k$ and $T$ are the Boltzmann constant and
temperature).
The Hamiltonian and the number operator of the interacting bosons are
denoted ${\bf H}$ and ${\bf N}$, respectively.
For interacting bosons this partition
function can be represented as a functional integral over
a complex scalar field $\Phi$.
In $D$ spatial dimensions this path
integral representation is given by
\cite{Feynman,Wiegel}
\begin{eqnarray}
Z(z,\beta) &=&
\int d[\Phi({\bf \xi}, \Theta),\Phi({\bf \xi}, \Theta)^*]
e^{- S(\Phi, \Phi^*)}
\label{path}
\end{eqnarray}
with the (dimensionless) action
\begin{eqnarray}
&&S(\Phi, \Phi^*) =
\int_0^{\beta/\beta_{\Lambda}} \int_{V\Lambda^D}
d\Theta d^D{\bf \xi}\{
\Phi^*({\bf {\xi}}, \Theta)
[-\frac{1}{2}\Delta_{\bf {\xi}} -
\nonumber\\
&&
M 
+
\frac{\partial}{\partial \Theta}]
\Phi({\bf {\xi}}, \Theta) +
\nonumber\\
&&
\frac{1}{2} 
\int_{V\Lambda^D}d^D{\bf \xi}'
|
\Phi({\bf {\xi}}, \Theta)|^2
[\beta_{\Lambda}V(\frac{{\bf {\xi}} - {\bf {\xi}}'}{\Lambda})]
| \Phi({\bf {\xi}'}, \Theta)|^2
\}.
\label{action}
\end{eqnarray}
The volume within which the Bose system is confined is denoted $V$.
For the sake of simplicity
this volume will be assumed
to be of cubic shape and the resulting spatial
confinement will
be described by imposing periodic boundary conditions
on the complex field $\Phi(\xi,\Theta)$.
In Eq.(\ref{path}) scaled quantities have been introduced
which involve a yet arbitrary
characteristic momentum $(\hbar \Lambda)$
whose
associated inverse temperature 
is given by $\beta_{\Lambda} = m/(\hbar^2 \Lambda^2)$.  
The mass of the interacting bosons
is denoted $m$.
The scaled chemical potential $M$ is
defined by $M=\mu\beta_{\Lambda}$.
The two-body 
interaction potential between the bosons is denoted
$V(\bf{x}-\bf{x}')$
and
$\xi={\bf x}\Lambda$ and $\Theta=\tau/(\hbar \beta_{\Lambda})$ are
scaled spatial coordinates and the scaled imaginary time.
The fugacity $z$ appearing in Eq.(\ref{action}) 
is related to the chemical potential $M$ by \cite{Wiegel}
\begin{equation}
{\rm ln} z = M - \frac{1}{2}\beta_{\Lambda}V({\bf x}={\bf 0}).
\end{equation}
The integration measure appearing in
Eq.(\ref{path}) is defined later (cf. Eq.(\ref{path1})).

Due to the trace-operation involved in the evaluation
of the partition function  of Eq.(\ref{partition})
the functional integration in Eq.(\ref{path}) has to be performed
over all complex fields $\Phi({\bf \xi}, \Theta)$ which are periodic
in the imaginary scaled time $\Theta$, i.e.
$\Phi({\bf \xi}, \Theta) = 
\Phi({\bf \xi}, \Theta + \beta/\beta_{\Lambda})$ \cite{Wiegel}.
If one also
imposes spatial periodic boundary conditions
which still preserve the translational invariance of the problem
it is convenient to transform
the functional integral of Eq.(\ref{path}) into the
momentum - frequency representation. This is achieved
by Fourier transforming the complex fields according to
\begin{eqnarray}
\Phi({\bf {\xi}},\Theta) = \sum_{{\bf m}, n}
[\frac{e^{i{\bf k}_{\bf m}\cdot {\bf {\xi}}}}{\sqrt{V\Lambda^D}}]
[\frac{e^{-i\omega_n \Theta}}{\sqrt{\beta/\beta_{\Lambda}}}]
\varphi_{{\bf m} n}
\label{Fourier}
\end{eqnarray}
with 
the Matsubara frequencies
$\omega_n = 2\pi n \beta_{\Lambda}/\beta$ 
and with the scaled wave vectors
${\bf k}^{(i)}_{\bf m} = 2\pi {\bf m}^{(i)}/(L_{i}\Lambda)$ (
${\bf m}^{(i)}$ being integer). Thereby the quantities
$L_{i}$ ($i=1,...,D$) denote the lengths of the
confining cubic volume.
Defining $(\hbar\Lambda)$ as the ultraviolet
momentum cut-off of the path integral,
which is much larger than all physically
significant momenta, 
implies that the scaled wave vectors
appearing in Eq.(\ref{Fourier}) are restricted to the region
$0\leq | {\bf k}_{\bf m} | \leq 1$.
Inserting this Fourier decomposition into Eq.(\ref{action}) one obtains
\begin{eqnarray}
&&S(\Phi,\Phi^*) =\sum_{{\bf m} n}
[-i\omega_n + \epsilon_{\bf m} -  M 
]
| \varphi_{{\bf m} n}|^2
+
\label{action1}
\\
&&
(V\Lambda^D)^{-1}(\beta/\beta_{\Lambda})^{-1}
\frac{1}{2} 
\sum_{{\bf m}_1 n_1;...;{\bf m}_4 n_4}
G({\bf k}_1 + {\bf k}_3)
(\varphi_{-{\bf m}_1 -n_1})^*
\times\nonumber
\\
&&
(\varphi_{-{\bf m}_2 -n_2})^*
\varphi_{{\bf m}_3 n_3}
\varphi_{{\bf m}_4 n_4}
\delta({\bf m}_1 + {\bf m}_2 + {\bf m}_3 + {\bf m}_4)
\times
\nonumber\\
&&
\delta(n_1 + n_2 + n_3 + n_4)
\end{eqnarray}
with
$\epsilon_{\bf m} = {\bf k}_{\bf m}^2/2$.
($\delta$ denotes the Kronecker-delta function.)
The two-particle interaction is characterized by its Fourier
components
\begin{equation}
G({\bf k}) = \int_{V\Lambda^D} d^D{\xi} e^{ik{\xi}}
\beta_{\Lambda} V(\xi/\Lambda).
\end{equation}
In the subsequent discussion we assume that the two-particle interaction
is of short range so that $G({\bf k})$ is independent of momentum, i.e.
$G({\bf k}) = G$.
Thus, introducing the momentum cut-off $(\hbar\Lambda)$
the path integral of Eq.(\ref{path})
becomes \cite{Wiegel,Wiegel1}
\begin{eqnarray}
Z(z,\beta) &=& [\prod_{{\bf m} n}\int
\frac{
d^2\varphi_{{\bf m} n}
}
{N_{{\bf m} n}}
]
e^{-S(\Phi,\Phi^*)}.
\label{path1}
\end{eqnarray}
The
normalization factors
\begin{eqnarray}
N_{{\bf m} n}
&=& \frac{\pi}{\zeta_{\bf m}
\beta_{\Lambda}/\beta
- i\omega_n}
\end{eqnarray}
with
$$2{\rm sinh}(\zeta_{\bf m}/2) =
e^{(\beta/\beta_{\Lambda})(\epsilon_{\bf m} - M)/2}$$
guarantee that in the limit of vanishing interactions, i.e.
$G\equiv 0$, the partition
function reduces to the well known expression for the ideal Bose gas
\cite{Wiegel}, i.e.
\begin{eqnarray}
Z(z,\beta)&=& \prod_{{\bf m}}\{
[
1 - e^{-(\beta/\beta_{\Lambda})(\epsilon_{\bf m} - M)}
]^{-1}
\}.
\label{ideal}
\end{eqnarray}
As expected on physical
grounds, for $\beta/\beta_{\Lambda},1/M \gg 1$ 
$Z(z,\beta)$
should become independent of the momentum cut-off $(\hbar \Lambda)$.
Obviously
this requirement is fulfilled for the ideal gas.
%

\subsection{Renormalization group approach for the evaluation of the
partition function}
One of the simplest methods of evaluating the partition function
of Eq.(\ref{path1})
is
the mean field approximation \cite{Wiegel1}. Thereby one expands
the field $\Phi(\xi,\Theta)$ appearing in Eq.(\ref{path})
quadratically around the
most probable, uniform
field configuration 
\begin{eqnarray}
\overline{\varphi}&=&
\sqrt{(V\Lambda^D)(\beta/\beta_{\Lambda}) M/G}
\label{most}
\end{eqnarray}
which is determined by the requirement\\
$\delta S(\Phi, \Phi^*)|_{\Phi=\Phi^*=
\overline{\varphi}/\sqrt{(V\Lambda^D)(\beta/\beta_{\Lambda})}}
= 0$.
The resulting Gaussian integrals of this quadratic expansion
can be performed in a straightforward way.
For $G>0$ 
this most probable field configuration is nonzero only if
$M > 0$.

A more sophisticated and more accurate method of evaluating
the partition function makes use of renormalization group methods.
The basic idea of the renormalization group procedure is to perform
the integrations in Eq.(\ref{path1}) successively \cite{Wilson}.
In each step only
field components $\delta\varphi_{{\bf m} n}$ are integrated out
whose momenta are located in
an infinitesimally small momentum shell around
the maximum momentum $(\hbar\Lambda)$, i.e. for which
$e^{-l} < | {\bf k}_{{\bf m}}| <1$ with $0<l\ll1$.
All other small-momentum field components
$\varphi_{{\bf m} n}$ outside this momentum shell,  which 
constitute the field $\phi_<$,
are left unchanged.
If only an infinitesimal momentum shell is integrated out,
a quadratic expansion of $S(\Phi, \Phi^*)$ with respect
to the fields 
$\delta\varphi_{{\bf m} n}$
is sufficient for the further evaluation of the partition function
\cite{Wegner}.

The desired aim of this integration over the large-momentum
field components is to obtain
a new scaled partition function
which is similar to the original
one except
for possible renormalizations of the characteristic parameters
$M$ and $G$
\cite{Kerson,Wilson}.
In particular this new scaled partition function has to
have the
same momentum cut-off $(\hbar \Lambda)$.
By repeated application of
this transformation 
one eventually integrates out all momentum components and obtains
the value of the partition function.
However, in general this desired
aim can be achieved only approximately.
Two commonly used approximation schemes involved
are perturbation theory
\cite{Wilson}
and the random phase approximation (RPA) \cite{Hertz}.
In our subsequent treatment the RPA
is employed. It is particularly useful in the case of a
non-vanishing most probable configuration
$\overline{\varphi}$ \cite{Stoof}.

In the subsequent development we separate
the
zero-temperature contribution
from
the grand thermodynamic potential according to
$\Omega(M,\beta) = -{\rm ln}Z(z,\beta)/(\beta/\beta_{\Lambda})$
\begin{equation}
\Omega(M,\beta) = \Omega(M,\beta\to\infty) + \omega(M,\beta).
\label{zero}
\end{equation}
For the zero-temperature contribution $\Omega(M,\beta\to\infty)$
accurate approximations are available.
In the simplest form of the mean field approximation
this zero-temperature contribution is given by \cite{Wiegel1}
\begin{equation}
\Omega(M,\beta\to\infty) = - (V\Lambda^D)
\frac{M^2}{2G}.
\label{meansimple}
\end{equation}
Note that the
depletion term \cite{Fetter} is not included in Eq.(\ref{meansimple}).
Typically this term
is negligibly small for a weakly interacting Bose gas.

It is shown in appendix A that
on integrating out momentum components of the fields
$\varphi_{{\bf m} n}$ which are located in the momentum shell
$e^{-l}\leq | {\bf k}_{\bf m}| \leq 1$ $(0<l\ll 1)$
the RPA yields the differential scaling relation
\begin{eqnarray}
\frac{d\omega(l)}{dl}&=&\frac{(V\Lambda^D)}{(\beta/\beta_{\Lambda})}
d(l) e^{-Dl} {\rm ln}(1 - e^{-\lambda(l)})
\label{omega}
\end{eqnarray}
with $\omega(M,\beta) \equiv \omega(l\to \infty)$.
Thereby the 
quantity
$\lambda(l) = (\beta(l)/\beta_{\Lambda})
\sqrt{\epsilon_>(\epsilon_> + 2M(l))}$ reflects the
Bogoliubov dispersion relation at each step of the renormalization
with
the scaled inverse temperature $\beta(l)=\beta e^{-2l}$.
The differential
scaling of the chemical potential $M(l)$ and of the
coupling strength $G(l)$ is governed by
the RG equations
\begin{eqnarray}
\frac{dM(l)}{dl} &=& 2M(l) +
\nonumber\\
&&
d(l)
(\beta(l)/\beta_{\Lambda})\frac{[{\rm coth}
(\lambda(l)/2) - 1]}{2\lambda(l)}\times\nonumber\\
&&
[2M(l) G(l) - G(l)2\epsilon_{>}] -\nonumber
\\
&&d(l)
(\beta(l)/\beta_{\Lambda})^3
\frac{M(l) G(l)}{2 \lambda(l)^2}\times\nonumber\\
&&\{
\frac{1}{2{\rm sinh}^2(\lambda(l)/2)} +\nonumber\\
&&
\frac{[{\rm coth}(\lambda(l)/2) - 1]}{\lambda(l)}
\}
[2\epsilon_{>} + M(l)]^2, 
\label{Mu}
\end{eqnarray}
\begin{eqnarray}
\frac{dG(l)}{dl} &=&
-(D - 2)G(l) +\nonumber
\\
&&
d(l)
(\beta(l)/\beta_{\Lambda})\frac{[{\rm coth}
(\lambda(l)/2) - 1]}{2\lambda(l)}
3G(l)^2 - \nonumber\\
&&
d(l)
(\beta(l)/\beta_{\Lambda})^3
\frac{G(l)^2}{2 \lambda(l)^2}\{
\frac{1}{2{\rm sinh}^2(\lambda(l)/2)} +\nonumber\\
&&
\frac{[{\rm coth}(\lambda(l)/2) - 1]}{\lambda(l)}
\}
[2\epsilon_{>} + M(l)]^2.
\label{Ge}
\end{eqnarray}
The (scaled) energy of the eliminated (infinitesimal) momentum shell
is denoted $\epsilon_> = 1/2$. 
The number of states per unit volume inside such an infinitesimal
momentum shell $e^{-l}\leq | {\bf k}_{\bf m}| \leq 1$
$(0<l\ll 1)$
is given by $[d(l)(\Lambda^D l)]$.
In the continuum limit, in $D$ spatial dimensions one obtains
the $l-$independent density
\begin{equation}
d(l) = 
\frac{\Omega_D}{(2\pi)^D}
\label{de}
\end{equation}
with
$\Omega_D = 2\pi^{D/2}/\Gamma(D/2)$ denoting the surface of a
D-dimensional sphere.
($\Gamma$ denotes the Euler-Gamma function.)

Eqs.(\ref{omega} - \ref{Ge}) 
constitute the set of differential
RG equations from which the 
temperature dependent part of the grand thermodynamic
potential $\omega(M,\beta) \equiv \omega(l\to\infty)$ can be evaluated.
They have to be solved subject to the initial conditions
\begin{eqnarray}
M(l=0) &=& M,\nonumber\\
G(l=0) &=& G,\nonumber\\
\omega(l=0) &=& 0.
\end{eqnarray}

The RG equations (15-17) are valid
for an arbitrary number of spatial dimensions $D$ of the interacting
Bose gas
as long as effects of confinement can be
described by periodic boundary conditions. 
Effects of confinement are characterized by
the quantity $d(l)$ which reduces to the result
of Eq.(\ref{de}) in the continuum limit, i.e. in the absence
of any confining influence.
If one or more degrees of freedom are confined, 
$d(l)$ is modified as one has to take into account
the resulting discrete energy levels in the confined
degree of freedom.
In a case where one degree of freedom of the interacting Bose
gas is confined, for example, one obtains
\begin{eqnarray}
d(l)&=&
\frac{\Omega_D}{(2\pi)^D}\frac{1}{2}\sum_{M=-\infty}^{\infty}
\int_{-1}^{1} dz e^{i z M\Lambda e^{-l} L_z} = \nonumber\\
&&
\frac{\Omega_D}{(2\pi)^D}
\frac{\pi + 2\pi[L_z e^{-l} \Lambda/(2\pi)]}{L_z e^{-l}\Lambda}.
\label{densityof}
\end{eqnarray}
Thereby $L_z$ is the length of the confining (square-well)
potential and $[x]$ denotes the largest integer which is smaller
or equal to $x$.

For dimensions less than or equal to two one has to reconsider
the validity of the ${\bf k}$-independent approximation
$G({\bf k}) = G$ of the interparticle potential.
If one assumes a spherically symmetric
short-ranged 
interparticle interaction then at low energies one may expand
$G({\bf k}) = G_0 + G_2{\bf k}^2+G_4{\bf k}^4+...$. 
It is straightforward  to demonstrate that for $l\to \infty$
(which implies $\beta(l)=\beta e^{-2l}\to 0$)
the trivial scaling of $G_k$ ($k=0,2,4,...$) is given by
$G_k(l)=G_k e^{[2-(D-2+k)]l}$. This implies that already
at $D=2$ not only
$G_0$ but also $G_2$ is no longer irrelevant. Thus for a reliable
description of the temperature dependence of two-dimensional
Bose gases also the ${\bf k}$-dependence of the
interparticle interaction has to be taken into account at least to
lowest order.

It is worth mentioning that
Eqs.(\ref{omega} - \ref{Ge}) reduce to
the corresponding mean field results
if one neglects fluctuations of the most probable configuration
$\overline{\varphi}$. Formally this is equivalent to
setting
$d(l) \equiv 0$ (or to putting 
$\Delta(\Phi_<) \equiv 0$ in Eq.(\ref{delta}) of appendix A).
In this case the scaling of the characteristic parameters $M(l)$
and $G(l)$ is
governed by the trivial scaling which depends on the
dimension $D$ of the system only.
Furthermore,  these mean field results also appear in the limit of
zero temperature where $\beta \to \infty$.

All thermodynamic properties which are derivable from the
grand thermodynamic potential $\Omega(M,\beta)$ can be determined
by these RG equations.
Thus  pressure $P$ and number of particles $N$, for example, are determined by the relations
\begin{eqnarray}
\frac{P\lambda_{th}^D}{kT} &=& -\Omega(M,\beta)\frac{\beta/\beta_{\Lambda}}{V\Lambda^D}
(2\pi\beta/\beta_{\Lambda})^{D/2},\\
\frac{N\lambda_{th}^D}{V} &=& -
(2\pi\beta/\beta_{\Lambda})^{D/2}\frac{\partial\Omega}{\partial M} (M,\beta)
\label{density}
\end{eqnarray}
with the thermal wave length $\lambda_{th}\equiv \sqrt{2\pi\hbar^2/(mkT)} =
\sqrt{2\pi\beta/\beta_{\Lambda}}/\Lambda$.
Another physical quantity which is of current experimental interest is the
spatially averaged second order coherence factor of the interacting
Bose gas. In terms of the
conjugate quantized field operators $\hat{\psi}(\bf{x})$
and $\hat{\psi}^{\dagger}(\bf{x})$ of the Bose gas it is defined by
\begin{eqnarray}
g^{(2)}(0) &=&\frac{\frac{1}{V}\int d^Dx
<\hat{\psi}^{\dagger}(\bf{x})\hat{\psi}^{\dagger}(\bf{x})
\hat{\psi}(\bf{x})\hat{\psi}(\bf{x})>}{
[\frac{1}{V}\int d^Dx<\hat{\psi}^{\dagger}(\bf{x})\hat{\psi}(\bf{x})>]^2}
\end{eqnarray}
with $< .  >$ denoting averaging over the thermodynamic
equilibrium distribution of the interacting Bose gas.
This definition is completely analogous to the corresponding quantity
in the quantum optical context.
As apparent from Eq.(1) this quantity
is related to the grand thermodynamic potential of Eq.(14) by
\begin{eqnarray}
g^{(2)}(0) &=& 2(V\Lambda^D)
\frac{\frac{\partial \Omega}{\partial G}(M,\beta,G)}
{[\frac{\partial \Omega}{\partial M}(M,\beta,G)]^2}.
\label{G2}
\end{eqnarray}
In the partial derivatives entering
Eq. (\ref{G2})
the grand thermodynamic potential $\Omega$ has to be considered as a function
of the fundamental parameters $(M,\beta,G)$. In order to keep the notation as simple
as possible this explicit
dependence on $G$ has not been indicated explicitly in all other previous equations.

Besides taking into account effects originating from confinement
the RG flow equations (15-17) differ from the ones derived previously
by Bijlsma and Stoof \cite{Stoof} also in two other respects. First of all,
separating off the zero-temperature contribution from the grand thermodynamic potential
according to Eq.(13) changes the properties of the flow equations. These changes
are particularly prominent 
in the region where $\lambda(l) \gg 1$, i.e. in the regions where
the value of the effective momentum cut-off is still large. 
The RG equations (15-17) have the property that for $\lambda(l)\gg 1$ the influence
of the nontrivial scaling which is proportional to the characteristic level
density $d(l)$ is exponentially small.
Secondly, evaluating the particle density directly from the
grand thermodynamic potential according to Eq.(\ref{density}) 
yields results which are slightly different from the ones
obtainable from the approximate
flow equation used in Ref.\cite{Stoof}.

\section{Numerical results}

In this section thermodynamic
properties of a confined interacting Bose
gas are investigated on the basis of the RG approach of Sec.II.
We concentrate
on cases in which one degree
of freedom of a three-dimensional Bose gas is confined
partly by changing the size of the confining
potential in one degree of freedom.
According to the theoretical developments of Sec.II
it is assumed that the resulting physical effects
of such a confinement can be described by periodic boundary conditions.
\centerline{\psfig{figure=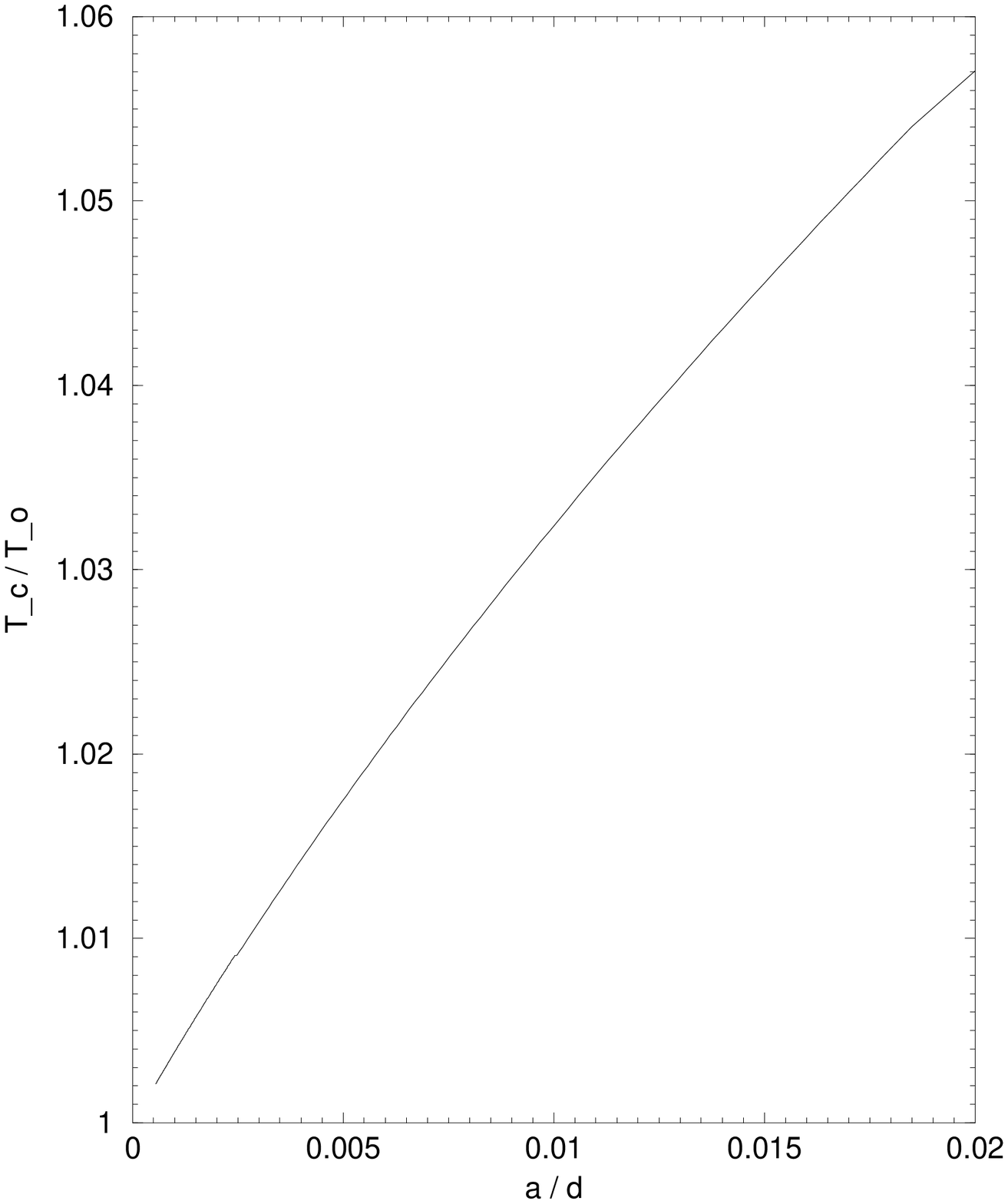,width=8.6cm,clip=}}
\begin{figure}
\caption{Dependence of the critical temperature $T_c$ of an interacting
Bose gas on the scattering length $a$. ($d = (V/N)^{1/3}$ denotes the mean distance
between atoms and $T_o$ is the critical temperature of an ideal Bose gas.)}
\label{fig1}
\end{figure}

Let us first of all concentrate on the critical properties of a homogeneous
interacting Bose gas in three spatial dimensions.
Though the corresponding characteristic critical exponents
are already well known \cite{Zinn},
non universal thermodynamic properties
of a  three-dimensional homogeneous Bose gas are still subject to
controversial discussions \cite{Holzmann1,Holzmann2,Huang2}.
One example of such a thermodynamic property
which is of topical interest is the critical temperature and its dependence on
the scattering length $a$ characterizing the strength of the interparticle
interaction. In three spatial dimensions this scattering length is related
to the characteristic interaction parameter $G$ of Sec. II by $G = 4\pi a\Lambda$.
The ratio between
the critical temperature $T_c$ of an interacting Bose gas and the
corresponding value of an ideal Bose gase $T_0$ can be determined easily from
Eq.(\ref{density}).
with the help of 
the relation $T_c/T_0 = [(N/V)\lambda_{th}^3]/2.612$ 
(Thereby the relation $(N/V)\lambda_{th}^3=\zeta(3/2)\approx 2.612$ has been used which applies
to an ideal homogeneous Bose gas at the transition temperature.)
The resulting predictions based on the RG approach of Sec. II are  depicted in Fig.1. 
One realizes that in this particular case of a repulsive interparticle interaction,
i.e. $a > 0$,
the critical temperature increases with increasing scattering
length. The quantitative dependence shown in Fig.1 is approximated
to a high degree of accuracy by the polynomial
\begin{eqnarray}
T_c &=& T_o [1.000+3.423(a/d)-29.986(a/d)^2 +\nonumber\\
&& 145.183(a/d)^3].
\label{approxim}
\end{eqnarray}
From this 
approximation we conclude that in the limit of a vanishing
interparticle interaction, i.e. $a\to 0$,
one obtains
$[(T_c - T_o)/T_o] = 3.423(a/d) + O[(a/d)^2]$.  
Eq.(\ref{approxim})
is in satisfactory agreement
with recent experimental results \cite{transition} and with the
perturbative and RG results of Refs.
\cite{Holzmann1,Holzmann2} and
\cite{Stoof}. 
\centerline{\psfig{figure=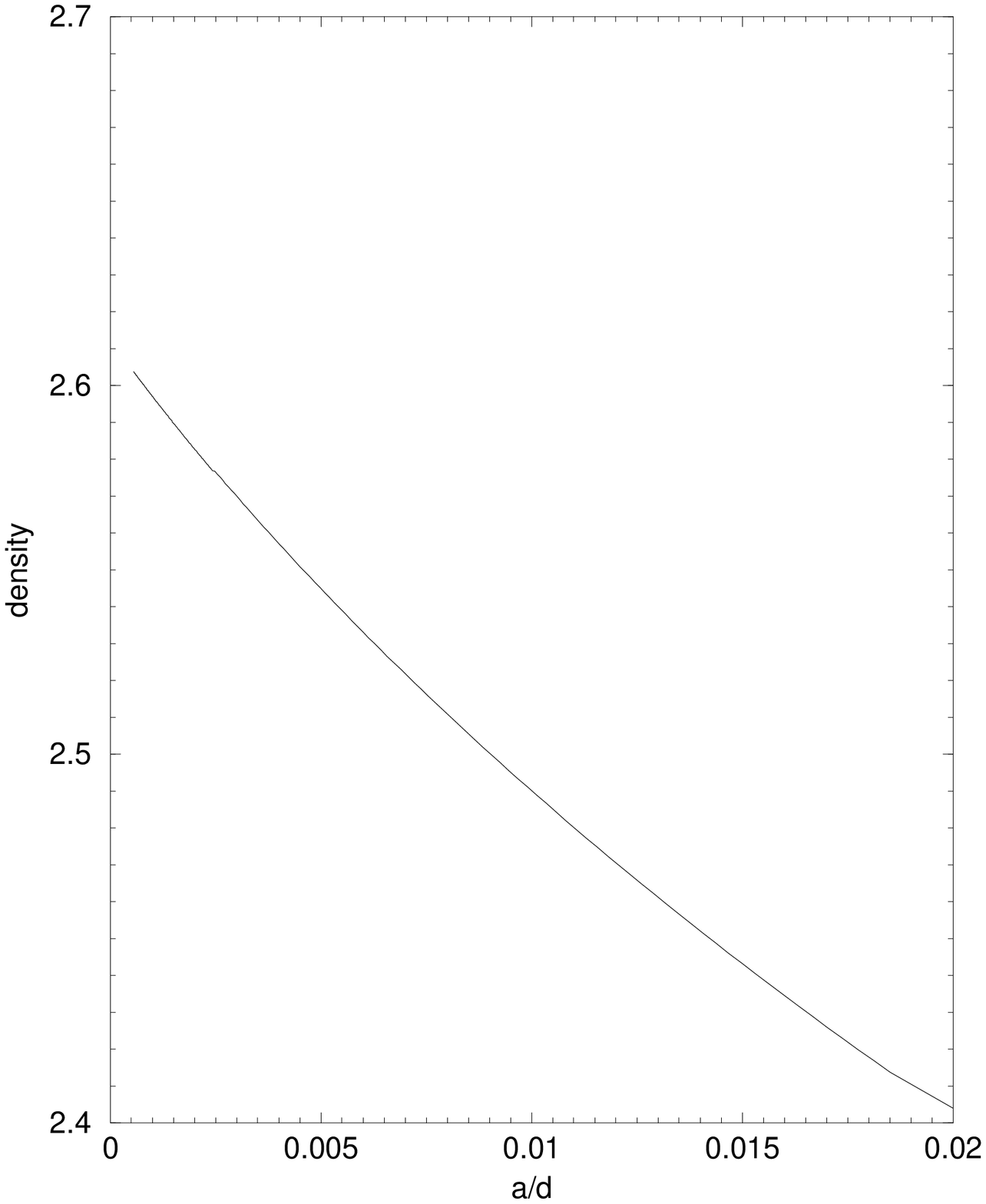,width=7.6cm,clip=}}
\begin{figure}
\caption{Dependence of the scaled critical density $[(N/V)\lambda_{th}^3]$
of an interacting Bose gas on the scattering length $a$. ($\lambda_{th}$ denotes
the thermal wave length.) }
\label{fig2}
\end{figure}

The critical density and its monotonically
decreasing dependence on the scattering length is shown in Fig.2.
\noindent
Note that in the limit $a\to 0$ one obtains the result
$(N/V)\lambda_{th}^3 \to 2.612$ as expected for an
ideal Bose gas.

The critical quantities depicted in
Figs.1 and 2  can be obtained from the RG approach
described in Sec. II in a straightforward way
from the knowledge of the critical trajectory. This critical trajectory
is the stable manifold of the unstable fixed point for $D=3$.
This unstable fixed point is given by 
$M_c=1/2$ and $g_c=\pi^2/2$ with $g(l)=G(l)/(\beta(l)/\beta_{\Lambda})$.
The associated eigenvalues of the linearized RG-equations for $M(l)$
and $g(l)$ are $\lambda_1 = 
(3-\sqrt{33})/6$
and $\lambda_2 =(3+\sqrt{33})/6$. The latter eigenvalue which
is associated with the unstable
manifold determines the various critical exponents of the
second order phase transition. Thus the scaling of the correlations length of the
interacting Bose gas, for example, is determined by the characteristic exponent
$\nu = 1/\lambda_2 = 0.686139$ which compares
well with the known result of $\nu = 0.67$ \cite{Zinn}
which is based on  a perturbative approach to the RG above the transition point.
\centerline{\psfig{figure=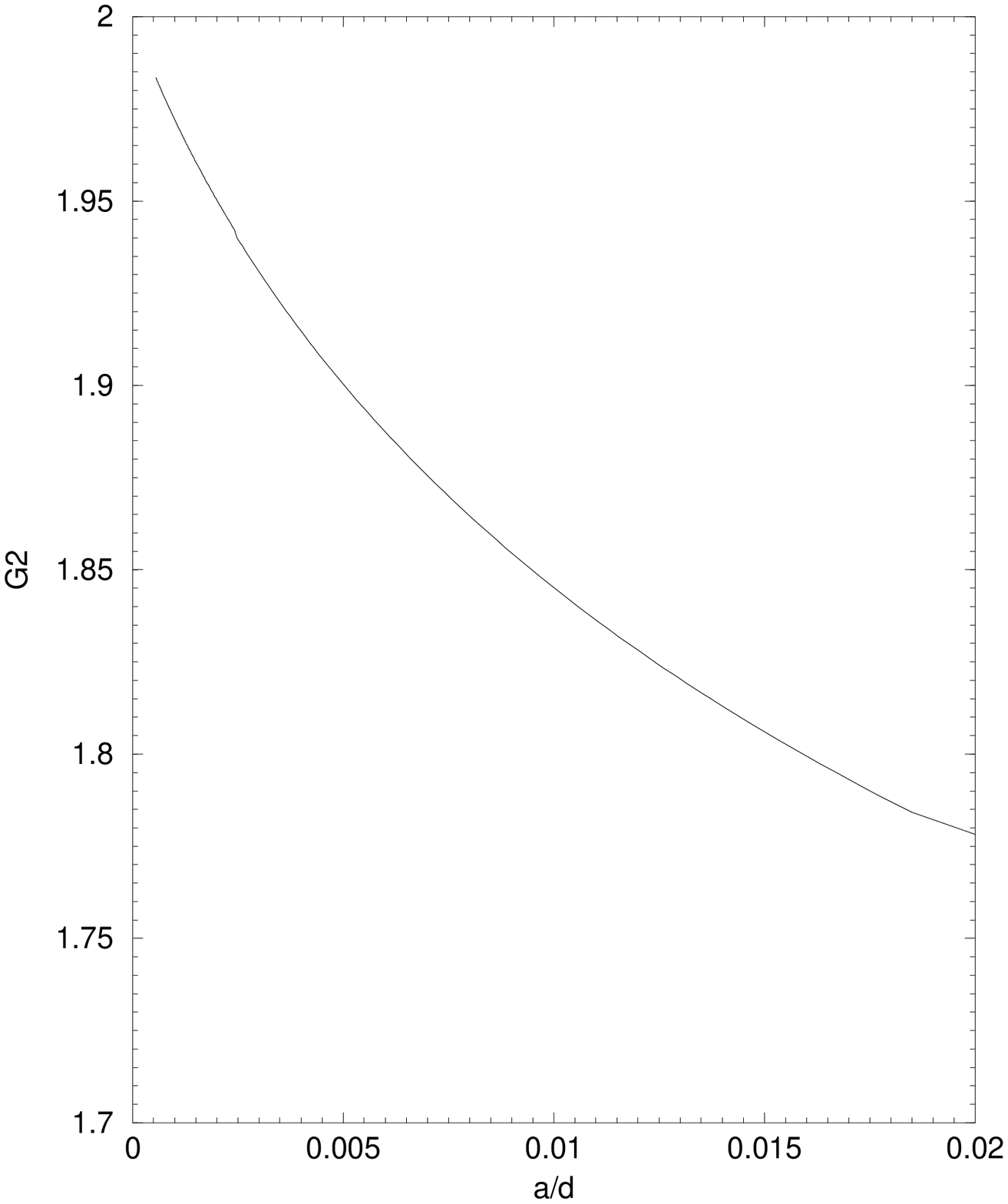,width=8.6cm,clip=}}
\begin{figure}
\caption{Dependence of the second order coherence factor 
$g^{(2)}(0)$ of an interacting Bose gas on the scattering length $a$.}
\label{fig3}
\end{figure}

One of the quantities which is accessible experimentally and  which has received
attention recently \cite{Ketterle} is the spatially averaged
second order coherence factor of an interacting Bose gas.
This quantity can be determined directly from the RG equations (15-17) of Sec. II
by using relation (\ref{G2}).
Its critical value at the transition point and its dependence on the scattering
length are depicted in Fig. 3.
For vanishingly small values of the interatomic interaction
this second order coherence factor approaches the value of two. This particular value is also known to
characterize photon bunching of a chaotic electromagnetic field.
 With increasing scattering length $a$
the critical value of the second order coherence
factor decreases thus indicating that with increasing
repulsive interactions the bosons tend to avoid each other.

Let us now investigate the influence of confining this homogeneous interacting
Bose gas with respect to one degree of freedom, say the z-direction. Furthermore,
let us assume 
that the influence of this confinement can be described
quantitatively by periodic boundary conditions.  In such a case the resulting
density of states $d(l)$ which enters the RG equations (15-17) is given by
Eq.(\ref{densityof}). In Fig. 4 isotherms of the  pressure of the interacting
Bose gas are depicted for various values of the characteristic confinement length
$L_z$ in the z-direction. 
Thereby above the phase transition, i.e. for negative values of the
scaled chemical potential $M(l)$, renormalization group equations
have been used which apply to
a vanishing order parameter \cite{Stoof}.

\centerline{\psfig{figure=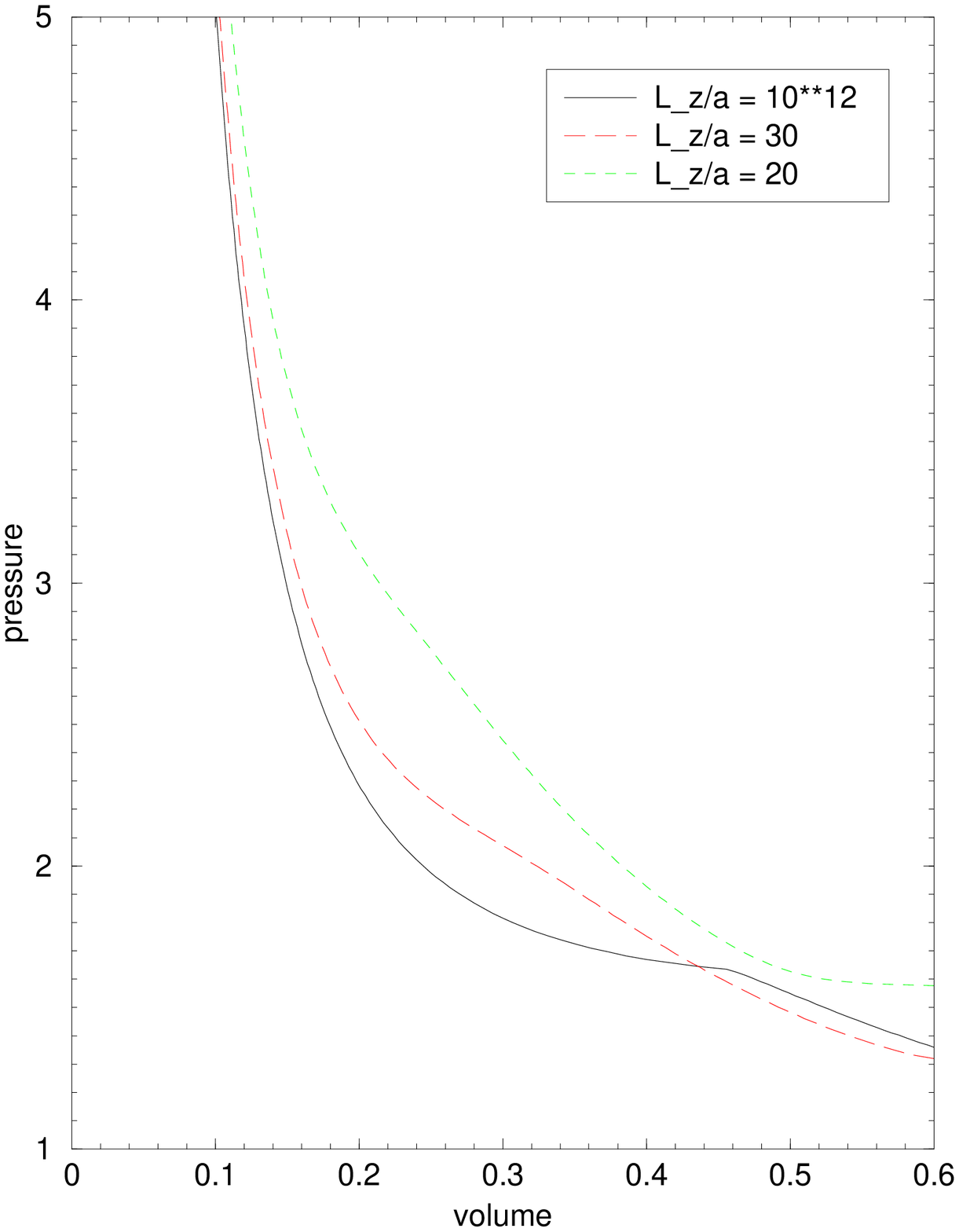,width=8.6cm,clip=}}
\begin{figure}
\caption{Isotherms of an interacting Bose gas for various values of the
characteristic length of confinement $L_z$. The scaled pressure 
$[p\lambda_{th}^3/(kT)]$ is depicted as a function of the scaled volume
$[(V/N)\lambda_{th}^{-3}]$.}
\label{fig4}
\end{figure}
\noindent
The temperature of these isotherms is chosen so that
$\lambda_{th}=10\sqrt{2\pi}a \approx 25 a$. For Rubidium atoms, for
example, with a scattering length of magnitude $a=5.3{\rm nm}$ 
this condition corresponds to a temperature of
$T = 1.98\times 10^{-6}{\rm K} $.
If the length of confinement $L_z$ is large in comparison with both
the thermal wave
length $\lambda_{th}$ and the scattering length $a$ (compare with full curve),
the characteristic signatures of a well pronounced second order phase transition
are realized at a critical volume of magnitude $(V/N)\lambda_{th}^{-3} = 0.456$
in agreement with the result of Fig. 2. 
As soon as the length of confinement $L_z$ becomes comparable to the thermal
wave length $\lambda_{th}$ the pressure dependence is modified significantly.
The most prominent feature of the depicted pressure dependence
is the disappearance of the characteristic signature of the second
order phase transition and the smoothing of this pressure dependence.

In view of the recent interest
in the behavior of the second order coherence
factors
of interacting Bose gases
its isothermal dependence on the
density of the interacting Bose gas is depicted in Fig. 5.
\centerline{\psfig{figure=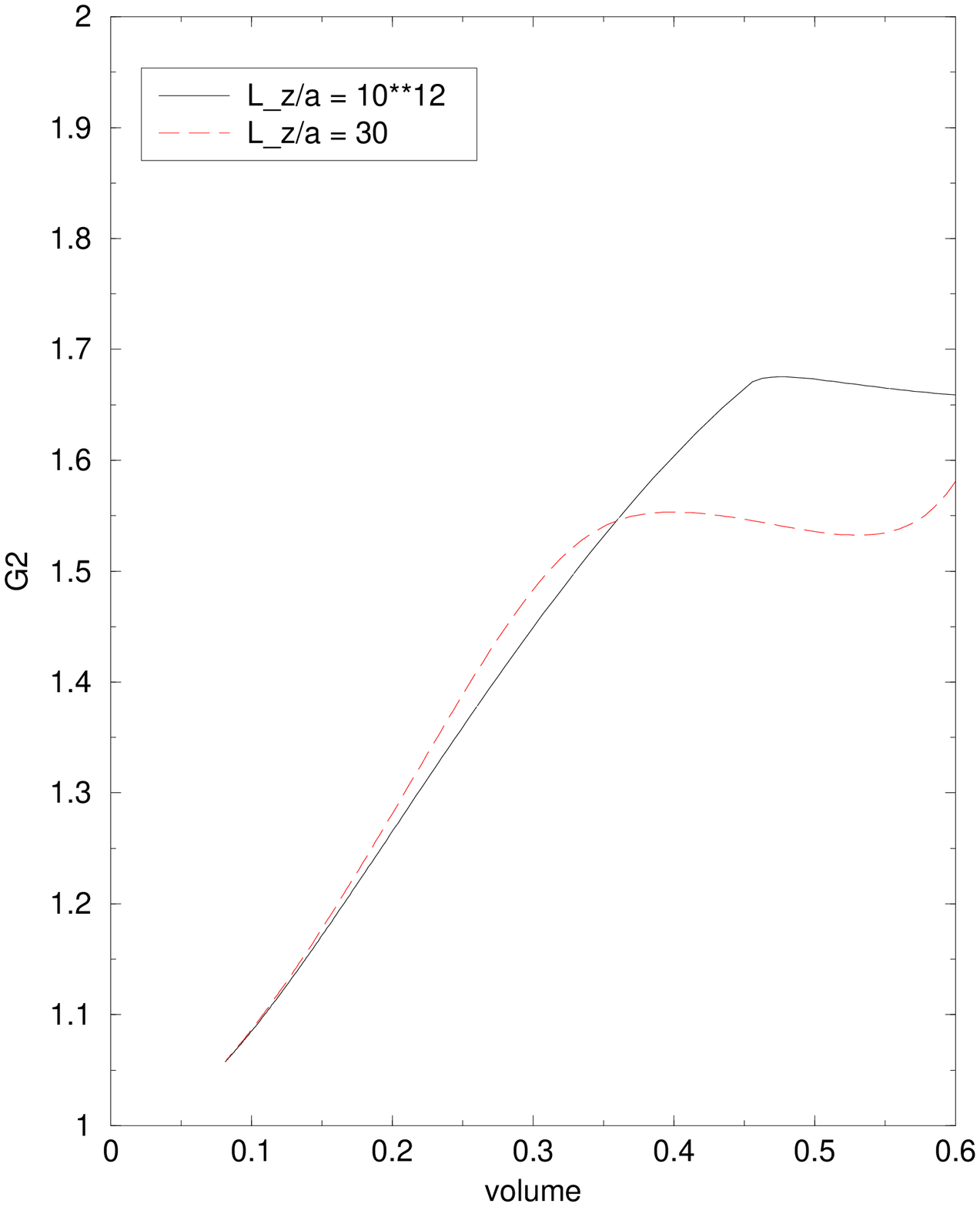,width=8.6cm,clip=}}
\begin{figure}
\caption{Isotherms of the second order coherence factor
$g^{(2)}(0)$
of an interacting Bose gas.}
\label{fig5}
\end{figure}
\noindent
The parameters chosen are the same as the ones used in Fig.4 .
Again for very large values of the confinement length $L_z$ one notices
a well pronounced second order phase transition
at the same critical volume
as in Fig.4. Far below the transition point, i.e. for very small densities,
the second order coherence
factor approaches a value of one which is the expected value for
a Bose condensate at zero temperature. 
Lowering the length of confinement $L_z$ in the
z-direction the characteristic
signature of the second order phase transition disappears.

section{Summary}
 
In summary, a theoretical description of thermodynamic properties
of an interacting confined Bose gas has been developed which is based on
a RG approach. Thereby physical effects which originate from the presence of 
confinement have been taken into account by periodic boundary
conditions.
Thus this approach yields a proper description of thermodynamic properties
also in cases in which the characteristic length
of confinement becomes comparable to the thermal wave length of a Bose gas
and in which local density approximations break down.
A more realistic and more complete treatment of boundary conditions 
will be postponed to subsequent work.
It has been demonstrated that this approach gives a reliable description of the
critical behavior of an interacting Bose gas which is consistent with recent
theoretical and experimental work. Furthermore, concentrating on the bunching properties of an
interacting Bose gas the influence of the confinement length on the
characteristic aspects of the phase transition has been worked out.

This work is supported by the Deutsche Forschungsgemeinschaft within
the Forschergruppe `Quantengase'.

\section*{Appendix A}

In this appendix the derivation of the RG transformation
of Sec. II is outlined in which an infinitesimal momentum shell of the 
partition function $Z(z,\beta)$ 
is integrated out. 
These equations are valid in the dynamical regime below the transition
temperature
which is characterized by a non-zero order parameter and a positive
renormalized chemical potential $M(l)$.

Starting point is the path integral representation 
of Eq.(\ref{path1}) with the (scaled) action
functional of Eq.({\ref{action1}).
Introducing a symmetry breaking, real-valued uniform field configuration
$\overline{\varphi}$ this action functional can be written in
the form
\begin{eqnarray}
&&S(\Phi, \Phi^*) =
\{-M \overline{\varphi}^2 + 
\frac{1}{2}g\overline{\varphi}^4\}+
[\varphi_{00}+\varphi_{00}^{*}][-M\overline{\varphi} +
g\overline{\varphi}^3] +\nonumber\\
&&
\sum_{{\bf m},n}[-i\omega_n + \epsilon_{\bf m} - M
+ 2
g\overline{\varphi}^2]
]| \varphi_{{\bf m} n}|^2
+\nonumber\\
&&
\sum_{{\bf m},n}
\frac{1}{2}
g\overline{\varphi}^2
[\varphi_{-{\bf m} -n}\varphi_{{\bf m} n}
+
\varphi_{-{\bf m} -n}^*\varphi_{{\bf m} n}^*
]+
\nonumber\\
&&
\sum_{{\bf m}_1,n_1... {\bf m}_3,n_3}
\overline{\varphi}g[
\varphi_{-{\bf m}_1 -n_1}^*\varphi_{{\bf m}_2 n_2}
\varphi_{{\bf m}_3 n_3} +\nonumber\\
&&
\varphi_{-{\bf m}_1 -n_1}^*\varphi_{-{\bf m}_2 -n_2}^*
\varphi_{{\bf m}_3 n_3}]\delta(n_1+n_2+n_3)
\delta({\bf m}_1+{\bf m}_2+{\bf m}_3)+
\nonumber\\
&&
\frac{1}{2}
\sum_{{\bf m}_1,n_1... {\bf m}_4 n_4}
g
\varphi_{-{\bf m}_1 -n_1}^*\varphi_{-{\bf m}_2 -n_2}^*
\varphi_{{\bf m}_3 n_3} 
\times
\nonumber\\
&&
\varphi_{{\bf m}_4 n_4}\delta(n_1+n_2+n_3+n_4)
\delta({\bf m}_1+{\bf m}_2+{\bf m}_3 +{\bf m}_4)
\label{action3}
\end{eqnarray}
with $g = 
(V\Lambda^D)^{-1}(\beta/\beta_{\Lambda})^{-1} G$.
In order to eliminate
the infinitesimal momentum shell around the maximum momentum
$(\hbar \Lambda)$, for which
$e^{-l} < | {\bf k}_{\bf m} | < 1$ $(0<l\ll 1)$,
one expands
the (scaled) action functional of Eq.(\ref{action1}) up to
second order in the large-momentum field components
$\delta \varphi_{{\bf m} n}$, i.e.
\begin{eqnarray}
S(\Phi,\Phi^*)&=&
S(\Phi_<,\Phi_<^*) +
\frac{1}{2}
\delta\varphi^T {\bf M}
\delta\varphi
\label{expand}
\end{eqnarray}
with
\begin{eqnarray}
\delta\varphi^T &=&
(...\delta\varphi_{{\bf m}n},
(\delta\varphi_{{\bf m}n})^*,
\delta\varphi_{-{\bf m} -n},
(\delta\varphi_{-{\bf m} -n})^*,...).
\end{eqnarray}
Thereby one chooses the symmetry breaking field $\overline{\varphi}$
identical to the most probable uniform field configuration
of Eq.(\ref{most}) so that
terms linear in $\varphi_{00}$ vanish
in Eq.(\ref{action3}).
The small-momentum field components $\varphi_{{\bf m} n}$
which are not integrated
out and which are kept constant constitute the field
$\Phi_<$.
The symmetric matrix ${\bf M}$ is given by
\begin{eqnarray}
{\bf M} &=&
\left(
\begin{array}{cccc}
. & . & . & . \\
0 &A + i\omega_n & B & 0 \\
A + i\omega_n & 0 & 0 & B^* \\
B & 0 & 0 & A - i\omega_n \\
0 & B^* & A - i\omega_n & 0 \\
. & . & . & . \\
\end{array}
\right)
\label{matrix}
\end{eqnarray}
with
\begin{eqnarray}
A&=&
\epsilon_> - M +
2g
[\overline{\varphi}^2
+
\overline{\varphi}\sum_{{\bf m} n}^{<}\varphi_{{\bf m} n}+
\nonumber\\
&&
\overline{\varphi}\sum_{{\bf m} n}^{<}(\varphi_{{\bf m} n})^*+
\sum_{{\bf m} n}^{<}(\varphi_{{\bf m} n})^*\varphi_{{\bf m} n}
],
\nonumber\\
B &=&
g
[\overline{\varphi}^2 
+2\overline{\varphi}
\sum_{{\bf m} n}^{<}(\varphi_{{\bf m} n})^*+
\sum_{{\bf m} n}^{<}(\varphi_{{\bf m} n})^*(\varphi_{-{\bf m} -n})^*
].
\label{AB2}
\end{eqnarray}
The quantity $\epsilon_> = 1/2$
denotes the scaled energy of the eliminated
momentum shell.
According to the RPA \cite{Hertz},
in Eq.(\ref{expand}) only products of large-momentum
field components $\delta\varphi_{{\bf m} n}$
are kept for which the sum of the
momenta and of the Matsubara frequencies vanishes.
Performing the Gaussian integrations over the fields
$\delta\varphi_{{\bf m} n}$ one 
obtains 
\begin{eqnarray}
Z (z,\beta) &=& [\prod_{{\bf m} n}^<\int
\frac{d^2\varphi_{{\bf m} n}}
{N_{{\bf m} n}}]
e^{-S_{new}(\Phi_<,\Phi_<^*)}
\label{out}
\end{eqnarray}
with the new effective action
\begin{eqnarray}
&&S_{new}(\Phi_<,\Phi_<^*) =
S(\Phi_<,\Phi_<^*)
-\nonumber\\
&& \frac{1}{2}\sum^{'}_{\bf m}
\sum_{n=-\infty}^{\infty}
{\rm ln}\frac{(\zeta_{{\bf m}n}\beta_{\Lambda}/\beta)^2 +
\omega_{n}^2}
{A^2 + \omega_n^2 - | B|^2}\nonumber\\
&=&
S(\Phi_<,\Phi_<^*)
- \frac{1}{2}\sum^{'}_{\bf m}
{\rm ln}
\frac{e^{(\beta/\beta_{\Lambda})(\epsilon_> - M)}}
{4{\rm sinh}^2[\lambda(\Phi_<)/2]}.
\label{effective1}
\end{eqnarray}
Thereby the relation
$${\rm sinh}^2(x)/x^2 =
\prod_{n=-\infty,n\neq 0}^{\infty}[1 + (\frac{x}{n\pi})^2]$$
has been used to perform the 
summation over all Matsubara frequencies.
The quantity $\lambda(\Phi_<)$ is given by
\begin{eqnarray}
\lambda(\Phi_<) &=&
(\beta/\beta_{\Lambda})\sqrt{\epsilon_{>}
[\epsilon_{>} 
+ 2M] -
\Delta(\Phi_<)}
\end{eqnarray}
with
\begin{eqnarray}
\Delta(\Phi_<) &=&
C\sum_{{\bf m}}^<\sum_{n=-\infty}^{\infty}
[
\varphi_{{\bf m} n} +
(\varphi_{{\bf m} n})^*] + 
\nonumber\\&&
D\sum_{{\bf m}}^<\sum_{n=-\infty}^{\infty}[
\varphi_{{\bf m} n} \varphi_{-{\bf m} -n}  +
(\varphi_{{\bf m} n})^* (\varphi_{-{\bf m} -n})^*] +
\nonumber\\
&&
(2D + C/\overline{\varphi})
\sum_{{\bf m}}^< \sum_{n=-\infty}^{\infty}|
\varphi_{{\bf m} n}|^2,\nonumber\\
C &=& (-4\epsilon_{>} - 2M)M/
\overline{\varphi}
,\nonumber\\
D &=& - 3M^2/\overline{\varphi}^2.
\label{delta}
\end{eqnarray}
The symbol $<$ in the summations of Eq.(\ref{delta})
indicates that only momenta have to be taken into account
which have not yet been integrated out.
The prime in the sum of Eq.(\ref{effective1})
indicates summation over all the eliminated momentum
components. In the continuum limit this latter summation
reduces to
\begin{eqnarray}
\sum_{\bf m}^{'} = (V\Lambda^D)\frac{\Omega_D}{(2\pi)^D}l
\hspace*{0.3cm}(0<l\ll 1).
\label{cont}
\end{eqnarray}

For the derivation of the RG transformation we
first of all split off 
the zero-temperature contribution according to
\begin{eqnarray}
S_{new}(\Phi_<,\Phi_<^*) &=& 
S(\Phi_<,\Phi_<^*)  +
(\beta/\beta_{\Lambda})\delta \Omega(M,\beta\to\infty) +
\nonumber\\
&&
\delta S(M,\beta).
\end{eqnarray}
The remaining temperature dependent contribution $\delta S(M,\beta)$
is then expanded
up to second order in the small-momentum field
components which constitute the field $\Phi_<$, i.e.
\begin{eqnarray}
\delta S(M,\beta)&=&
-
\sum_{\bf m}' \frac{1}{2}
{\rm ln}\frac{e^{\lambda}}
{4 {\rm sinh}^2(\lambda/2)} + \nonumber\\
&&
\frac{1}{2}
\frac{(\beta/\beta_{\Lambda})^2\Delta(\Phi_<)}{2\lambda}
\sum^{'}_{\bf m}
\frac{d}{d\lambda}{\rm ln}\frac{e^{\lambda}}
{4{\rm sinh}^2(\lambda/2)}-\nonumber\\
&&
\frac{1}{2}
[\frac{(\beta/\beta_{\Lambda})^2\Delta(\Phi_<)}{2\lambda}]^2
\sum^{'}_{\bf m}
\frac{1}{2}\frac{d^2}{d\lambda^2}{\rm ln}\frac{e^{\lambda}}
{4{\rm sinh}^2(\lambda/2)}
\label{effective2}
\end{eqnarray}
with $\lambda = \lambda(\Phi_<\equiv 0)$ and with
$\Delta(\Phi_<)$ as defined by Eq.(\ref{delta}).
Note that the terms involving $e^\lambda$ result from the 
separation of the zero-temperature contribution according to
Eq.(\ref{zero}).
This expanded effective action involves terms linear in the
field $\Phi_<$. These linear terms can be
absorbed in a redefinition of
the most probable configuration, i.e.
\begin{eqnarray}
\overline{\varphi}_{new}&=&
\overline{\varphi} + \Delta\overline{\varphi}
\label{phinew}
\end{eqnarray}
with
\begin{eqnarray}
\Delta\overline{\varphi} &=&
\frac{1}{4M}\sum^{'}_{\bf m}
(\beta/\beta_{\Lambda})^2 
\frac{1}{2\lambda}[
{\rm coth} 
(\lambda/2) -1]C.
\label{Dphi}
\end{eqnarray}
Eq.(\ref{effective2})
implies a change of
the chemical potential, i.e. 
\begin{eqnarray}
M_{new} &=&
M + \Delta M 
\label{Mnew}
\end{eqnarray}
with
\begin{eqnarray}
&&\Delta M =-
\sum^{'}_{\bf m}
(\beta/\beta_{\Lambda})^2
\frac{1}{2\lambda}
[{\rm coth} 
(\lambda/2) - 1]
D
+ 
2\frac{\Delta\overline{\varphi}}
{\overline{\varphi}}
M
-\nonumber\\
&&
\sum^{'}_{\bf m}
\frac{1}{2}
(\beta/\beta_{\Lambda})^4
\frac{1}{4\lambda^2}
\{
\frac{1}{2{\rm sinh}^2(\lambda/2)}
+
\frac{1}{\lambda}
[{\rm coth} 
(\lambda/2) - 1]
\}
C^2.
\label{DM}
\end{eqnarray}
The quantities $C$ and $D$ are defined by Eq.(\ref{delta}).
The corresponding change of the scaled
two-body coupling strength $G$ is 
determined by the requirement that 
\begin{eqnarray}
\overline{\varphi}_{new} &=&
\sqrt{(V\Lambda^D)(\beta/\beta_{\Lambda}) M_{new}/G_{new}}
\end{eqnarray}
in accordance with Eq.(\ref{most}).
This implies that
\begin{eqnarray}
G_{new} &=&
G + \Delta G 
\label{Gnew}
\end{eqnarray}
with
\begin{eqnarray}
\Delta G &=&G[\frac{\Delta M}{M} -
2\frac{\Delta \overline{\varphi}}{\overline{\varphi}}].
\label{DG}
\end{eqnarray}
Eqs.(\ref{phinew} - \ref{DM})
and (\ref{Gnew} - \ref{DG}) characterize the
Kadanoff transformation \cite{Kerson}, i.e. the
elimination of the infinitesimal momentum shell for which
$e^{-l} < | {\bf k_m}| <1$ with $l \ll 1$.
The RG transformation 
is completed by performing the
scaling transformations which restore the original momentum cut-off
$(\hbar \Lambda)$, namely
\begin{eqnarray}
{\bf k}_{\bf m'}
&=&
{\bf k}_{\bf m}e^{l},\nonumber\\
\omega_{n'}&=&
\omega_{n}e^{2l},\nonumber\\
V' &=& V e^{-Dl},\nonumber\\
\beta' &=& \beta e^{-2l},\nonumber\\
\varphi^{'}_{n'\bf m'}
&=&
\varphi_{n\bf m}e^{-l}.
\end{eqnarray}
With
$M(l)=(M+\Delta M)e^{2l}$, 
$G(l)=(G+\Delta G)e^{(2-D)l}$, 
$\beta(l)=\beta e^{-2l}$ 
these scaling relations together with the Kadanoff transformations
of Eqs.(\ref{phinew} - \ref{DM},\ref{Gnew} - \ref{DG}) imply
the RG equations of Eqs.(\ref{omega} - \ref{Ge}).

\section*{Appendix B}

For the sake of completeness
in this appendix the RG equations are summarized which apply to
the case of a vanishing order parameter above the transition point.

In this dynamical regime
the scaled chemical potential $M(l)$ might become negative.
The RG equations which apply to this regime
of negative values of $M(l)$
can be obtained in a straightforward way
with the help of second order perturbation theory by assuming a
vanishing order parameter
\cite{Stoof,Hertz}.
Thus, for $M(l) < 0$
one obtains the RG flow equations
\begin{eqnarray}
&&\frac{d\omega(l)}{dl} = \frac{(V\Lambda^D)}{(\beta/\beta_{\Lambda})}
d(l) e^{-Dl} {\rm ln}(1 -
e^{-(\beta/\beta_{\Lambda})(\epsilon_> - M(l))}),
\nonumber\\
&&\frac{dM(l)}{dl} = 2M(l) -d(l) G(l) \{\frac{1}
{{\rm tanh}[(\beta/\beta_{\Lambda})(\epsilon_> -
M(l))/2]}
- 1\},
\nonumber\\
&&\frac{dG(l)}{dl} = (2 - D)G(l) -d(l) G^2(l)\times\nonumber\\
&& \{
\frac{1}{{\rm tanh}[(\beta/\beta_{\Lambda})(\epsilon_> - M(l))/2]}
\frac{1}{2(\beta/\beta_{\Lambda})^2(\epsilon_> - M(l))}
+\nonumber\\
&&(\beta/\beta_{\Lambda})\frac{1}{{\rm sinh}^2
[(\beta/\beta_{\Lambda})(\epsilon_> - M(l))/2]}
\}.
\label{RG2}
\end{eqnarray}
Starting from a positive chemical potential,
i.e. $M(l=0) \equiv M >0$,
one has to switch from Eqs. (16-17) to
Eq. (\ref{RG2}) as soon as $M(l)$ becomes negative in the
process of renormalization.
If one starts from a negative chemical potential, i.e.
$M(l=0)\equiv M < 0$, one has to solve Eq. (\ref{RG2}) with the
intial conditions
\begin{eqnarray}
\omega (l=0) &=&0,\nonumber\\
M(l=0) &=& M <0,\nonumber\\
G(l=0) &=& G.
\end{eqnarray}
The corresponding grand thermodynamic potential is given by
$\Omega(M,\beta) \equiv \omega(l\to \infty)$.


\end{document}